%% file: draft.tex
\RequirePackage{lineno}
\documentclass[twocolumn,aps,tightenlinefs,superscriptaddress,preprintnumbers,nofootinbib]{revtex4}
\usepackage{graphicx,color,dcolumn,booktabs,bm}
\usepackage{longtable,lscape}
\usepackage{amsmath}
\usepackage{indentfirst}
\usepackage{epsfig}
\usepackage{feynmf}   
\usepackage{epstopdf}   
\usepackage{slashed}  
\usepackage{cases}
\usepackage{color}
\usepackage{multirow}
\usepackage{ulem}
\usepackage{verbatim}
\usepackage[colorlinks,linkcolor=red,anchorcolor=green,citecolor=blue]{hyperref}
\usepackage{mathrsfs}
\usepackage{rotating}
\usepackage{threeparttable}
\usepackage{lineno}
\usepackage{subfigure}
\usepackage{soul}
\usepackage{orcidlink}

\graphicspath{{ps}}

\newcommand{\GG}{\gamma\gamma}
\newcommand{\lamc}{\Lambda_c^+}
\newcommand{\pkpi}{pK^-\pi^+}
\newcommand{\pks}{pK^0_S}
\newcommand{\lceta}{\Lambda_c^+\eta}
\newcommand{\pdz}{pD^0}
\newcommand{\sgmpi}{\Sigma_c(2455)\pi}

\newcommand{\tripi}{\pi^0\pi^+\pi^-}
\newcommand{\konepi}{K^-\pi^+}
\newcommand{\ktwopi}{K^-\pi^+\pi^0}
\newcommand{\ktripi}{K^-\pi^+\pi^+\pi^-}
\newcommand{\BF}{\mathcal{B}}

\begin{document}



\title{\quad\\[0.1cm] \boldmath Search for charmed baryons in the $\Lambda_c^+\eta$ system and measurement of the branching fractions of $\Lambda_c(2880)^+$ and $\Lambda_c(2940)^+$ decaying to $\Lambda_c^+\eta$ and $pD^0$ relative to $\Sigma_c(2455)\pi$}

\include{pub676-orcid.tex}

\begin{abstract}
     We search for excited charmed baryons in the $\lceta$ system using a data sample corresponding to an integrated luminosity of 980 $\rm fb^{-1}$. The data were collected by the Belle detector at the KEKB $e^{+}$$e^{-}$ asymmetric-energy collider. No significant signals are found in the $\lceta$ mass spectrum, including the known $\Lambda_c(2880)^+$ and $\Lambda_c(2940)^+$. Clear $\Lambda_c(2880)^+$ and $\Lambda_c(2940)^+$ signals are observed in the $\pdz$ mass spectrum. 
     We set upper limits at 90\% credibility level on ratios of branching fractions of $\Lambda_c(2880)^+$ and $\Lambda_c(2940)^+$ decaying to $\lceta$ relative to $\Sigma_c(2455)\pi$ of $<0.13$ for the $\Lambda_c(2880)^+$ and $<1.11$ for the $\Lambda_c(2940)^+$.  
     We measure ratios of branching fractions of $\Lambda_c(2880)^+$ and $\Lambda_c(2940)^+$ decaying to $\pdz$ relative to $\Sigma_c(2455)\pi$ of $0.75 \pm 0.03(\text{stat.}) \pm 0.07(\text{syst.})$ for the $\Lambda_c(2880)^+$ and $3.59 \pm 0.21(\text{stat.}) \pm 0.56(\text{syst.})$ for the $\Lambda_c(2940)^+$.
     
\end{abstract}

\maketitle

\tighten

\section{\boldmath Introduction}
\setstcolor{red}
     Charmed baryons provide a good opportunity to study the dynamics of quark confinement~\cite{intro-theo1,intro-theo2,intro-theo3,intro-theo4,intro-theo5}. Usually, singly-charmed baryons are considered to be bound states of a charm quark and a light diquark. In this model, the Jacobi coordinate $\rho$ is used to label the degree of freedom between the light quarks and $\lambda$ to label the degree of freedom between the charm quark and the center of mass of the diquark. The excitation of the diquark is called $\rho$-mode excitation, while the excitation between the charm quark and the diquark is called $\lambda$-mode excitation. The excitation energy for the $\rho$-mode is expected to be higher than the one for $\lambda$-mode by a factor of $\sqrt{3}$ in the heavy-quark limit~\cite{intro-qm}. 

     In the $\Lambda_c^+$ sector, both the $\Lambda_c(2880)^+$ and the $\Lambda_c(2940)^+$ have been observed in several decay modes. The $\Lambda_c(2880)^+$ was first observed by CLEO in the $\Lambda_c^+\pi^+\pi^-$ decay mode, where some of the events resonated through $\Sigma_c(2455) \pi$~\cite{intro-lcpipi}, and later reported by BaBar in the $\pdz$ mass spectrum~\cite{intro-pdz}. The $\Lambda_c(2940)^+$ was first seen by BaBar in the $\pdz$ decay mode~\cite{intro-pdz} and confirmed by Belle in the $\Sigma_c(2455)\pi$ decay~\cite{intro-sgmpi}. The spin-parities of the two states were also studied. Belle analyzed the $\Sigma_c(2455/2520)\pi$ decay modes and measured the spin of the $\Lambda_c(2880)^+$ to be $5/2$~\cite{intro-sgmpi}. LHCb studied the $\Lambda_b^0 \to D^0p\pi^-$ decay and measured the favored $J^P$ of $\Lambda_c(2940)^+$ to be $3/2^-$, but they could not exclude spins from $1/2$ to $7/2$~\cite{intro-lc2940-jp}. The inner structure of the $\Lambda_c(2940)^+$ remains a puzzle.  The $\Lambda_c(2940)^+$ may be a conventional excited charmed baryon, but has also been explained as an $ND^*$ molecular state with potential spin-parity assignments including $1/2^+$, $1/2^-$, and $3/2^-$~\cite{intro-ND1, intro-ND2, intro-ND3}.

     $\Lambda_c^+\eta$ is a good channel to search for excited $\Lambda_c^+$ baryons. Any signal in $\lceta$ is likely to be an excited $\Lambda_c^+$ rather than a $\Sigma_c^+$, as for the latter decays to $\Lambda_c^+ \pi$ are allowed by isospin and are likely to dominate. Furthermore, studies of the $\Lambda \eta$ channel in the strange-baryon sector suggest $\Lambda_c^+\eta$ to be an interesting channel.
     In addition to excited $\Lambda$ states decaying into $\Lambda\eta$, such as the $\Lambda(2000)$~\cite{intro-lbd2000}, a narrow enhancement was observed recently in the $pK^-$ channel near the $\Lambda\eta$ threshold, which was identified as a threshold cusp~\cite{intro-ld}. 

     Much attention has been paid to the $pD^0$ system as an analogue to the $NK$ system, where there exists the $\Lambda(1405)$ resonance as a $NK$ quasi-bound state near threshold~\cite{intro-kn1,intro-kn2,intro-kn3}. Thus the $pD^0$ system can be studied from the viewpoint of the description of the $\Lambda_c$ charmed baryons. BaBar has reported $\Lambda_c(2880)^+$ and $\Lambda_c(2940)^+$ in the $\pdz$ mass spectrum~\cite{intro-pdz}, while Belle has not previously investigated the $pD^0$ spectrum in direct $e^+e^-$ annihilation.

     In this study, which is based on the full Belle dataset, 
     we study for the first time the $\Lambda_c^+\eta$ system and search for singly-charmed baryons, including $\Lambda_c(2880)^+$, $\Lambda_c(2940)^+$ and other yet unknown states, in a mass region from 2.83 to 3.15 GeV/$c^2$. In addition, we measure ratios of branching fractions of $\Lambda_c(2880)^+$ and $\Lambda_c(2940)^+$ decaying to $\lceta$ and $\pdz$ relative to the reference mode, $\Sigma_c(2455)\pi$, where $\Sigma_c(2455)\pi$ is the combination of $\Sigma_c(2455)^0 \pi^+$ and $\Sigma_c(2455)^{++} \pi^-$. 

\section{\boldmath data sample and belle detector}
\label{sec:sample}

     We perform our analysis using a data sample corresponding to an integrated luminosity of 980 $\rm fb^{-1}$, collected with the Belle detector at the KEKB asymmetric-energy $e^+e^-$ collider~\cite{KEKB1,KEKB2}. Most of the data were recorded at the $\Upsilon(4S)$ resonance, the rest was collected at other $\Upsilon(nS)$ states, with $n$ = 1,~2,~3, or 5, or at 60 MeV below the $\Upsilon(4S)$. 

    The Belle detector is a large-solid-angle magnetic spectrometer that consists of a silicon vertex detector (SVD), a 50-layer central drift chamber (CDC), an array of aerogel threshold Cherenkov counters (ACC), a barrel-like arrangement of time-of-flight scintillation counters (TOF), and an electromagnetic calorimeter (ECL) comprised of CsI(Tl) crystals. All these detector components are located inside a superconducting solenoid coil that provides a 1.5~T magnetic field. An iron flux-return located outside of the coil is instrumented to detect $K_L^0$ mesons and to identify muons. The detector is described in detail in Refs.~\cite{Belle1,Belle2}.

     We use samples of simulated $e^+e^- \to c\bar{c}$ Monte Carlo (MC) events for the optimization of selection criteria, the estimation of background contributions, and the determination of signal detection efficiencies. MC samples are generated by {\sc EvtGen}~\cite{evtgen} and then propagated through a detector simulation based on {\sc GEANT3}~\cite{geant3}. 
     The package {\sc PYTHIA}~\cite{pythia} is used to simulate $e^+e^- \to c\bar{c} \to \Lambda_c^{*+} X$ signal events, where $\Lambda_c^{*+}$ stands for $\Lambda_c(2880)^+$, $\Lambda_c(2940)^+$, or one of possible new excited $\Lambda_c$ states, and $X$ denotes anything. The masses and widths of $\Lambda_c(2880)^+$ and $\Lambda_c(2940)^+$ are fixed to the world-average values~\cite{pdg}. The masses of the unknown $\Lambda_c$s are set to values from 2.850 to 3.125 GeV/$c^2$ in steps of 25 MeV/$c^2$, and for each mass value the widths are set to 0, 10, 20, 30, and 40 MeV. We generate the $\Lambda_c^{*+} \to \lceta$ and $\Lambda_c(2880)^+/\Lambda_c(2940)^+ \to \pdz/\sgmpi$ samples with a phase-space model~\cite{phsp} and also the secondary decays $\lamc \to \pkpi/\pks$, $\eta \to \GG$, $D^0 \to \konepi/\ktwopi/\ktripi$, and $\Sigma_c(2455) \to \lamc\pi$ are all generated with a phase-space model. The $\eta \to \tripi$ {decay} is generated with its known substructures~\cite{pdg}. We take the effect of final-state radiation from charged particles into account with the {\sc PHOTOS} package~\cite{photons}. We use the MC samples with zero width for $\Lambda_c^{*+}$ in order to estimate the intrinsic mass resolution. Generic MC samples of $\Upsilon(1S,~2S,~3S)$ decays, $\Upsilon(4S)\to B^{+}B^{-}/B^{0}\bar{B}^{0}$, $\Upsilon(5S) \to B_{s}^{(*)}\bar{B}_{s}^{(*)}/B^{(*)}\bar{B}^{(*)}(\pi)/\Upsilon(4S)\gamma$, and $e^+e^- \to q\bar{q}$ $(q=u,~d,~s,~c)$ at $\sqrt{s}$ = 10.52, 10.58, and 10.867~GeV, corresponding to two times the integrated luminosity of the data, are used to optimize selection criteria and perform background studies.

\section{\boldmath Event selection}
\label{sec:selection}
     We study the $\Lambda_c^{*+}$ in three decay modes: $\lceta$, $\pdz$, and $\Sigma_c(2455)\pi$, with $\Lambda_c^+ \to \pkpi$/$\pks$, $\eta \to \GG$/$\tripi$, $D^0 \to \konepi$/$\ktwopi$/$\ktripi$, and $\Sigma_c(2455) \to \Lambda_c^+\pi$. Event selections are optimized by maximizing a figure-of-merit defined as $\epsilon/\sqrt{B}$, where $\epsilon$ denotes the detection efficiency of the signal process and $B$ is the number of background events obtained from generic MC samples in a mass region from 2.870 to 2.895 GeV/$c^2$, which is defined as the $\Lambda_c(2880)^+$ signal region.

     Final-state charged particles, $\pi^{\pm}$, $K^{\pm}$, and $p(\bar{p})$, are identified using information from the tracking systems (SVD, CDC) and the particle identification detectors (CDC, ACC, TOF), which is combined into the likelihood ratio $\mathcal{L}(A|B) =\mathcal{L}(A)/[\mathcal{L}(A) + \mathcal{L}(B)]$, where $A$ and $B$ are $\pi$, $K$, or $p$ as appropriate~\cite{pidcode}. We require protons to have $\mathcal{L}(p|K)>0.6$ and $\mathcal{L}(p|\pi)>0.6$, kaons to have $\mathcal{L}(K|p)>0.6$ and $\mathcal{L}(K|\pi)>0.6$, and pions to have $\mathcal{L}(\pi|p)>0.6$ and $\mathcal{L}(\pi|K)>0.6$. A requirement of $\mathcal{L}(e|\rm hadrons)=\mathcal{L}(e)/[\mathcal{L}(e)+\mathcal{L}(\rm hadrons)]<0.95$ is applied for all charged particles to suppress electrons, where $\mathcal{L}(e)$ and $\mathcal{L}(\rm hadrons)$ are obtained from the ECL in addition to tracking and particle identification systems~\cite{eidcode}. The momentum-averaged identification efficiencies of protons, kaons, and pions are 90\%, 90\%, and 93\%, respectively. 

     We also require the distance of the closest approach of all charged-particle tracks with respect to the interaction point to be less than 2.0 cm along the direction opposite to the $e^+$ beam and to be less than 0.2 cm in the transverse plane. In addition, invariant masses of pion, kaon, and proton pairs with the same sign are required to be larger than 0.28 GeV/$c^2$, 0.989 GeV/$c^2$, and 1.878 GeV/$c^2$, respectively, in order to reject duplicated tracks. 

     A cluster detected in the ECL not matching any tracks is identified as a photon candidate. Candidates for $\pi^0 \to \GG$ {are} selected from {photon pairs, where each photon has an} energy greater than 0.05 GeV, {and are} discarded once the invariant mass of the two photons {lies} outside the range from 0.12 GeV/$c^2$ to 0.15 GeV/$c^2$. {We reconstruct the $\eta$ candidates using two modes: $\GG$ and $\tripi$}. {Candidates for} $\eta \to \GG$ are reconstructed from {photon pairs} whose invariant mass satisfies 0.52 GeV/$c^2$ $<M(\GG)<$ 0.57 GeV/$c^2$. The energies of both photons are required to be greater than 0.3 GeV. {We select $\eta \to \tripi$ candidates} with $\pi^0 \to \GG$ {by requiring} 0.540 GeV/$c^2$ $<M(\tripi)<$ 0.555 GeV/$c^2$, where $M(\tripi)$ is the invariant mass of $\tripi$. {To suppress background events we require in addition that} the photons from the $\eta \to \pi^+\pi^-\pi^0$, $\pi^0 \to \gamma\gamma$ decay have an energy greater than 0.1 GeV. 
     We adjust the four-momenta of the daughter particles within their uncertainties by constraining the invariant masses of the systems of final-state particles to the nominal $\pi^0$ and $\eta$ masses~\cite{pdg}, respectively.  

\begin{figure}[htbp]
        \begin{center}
        \includegraphics[width=0.44\textwidth]{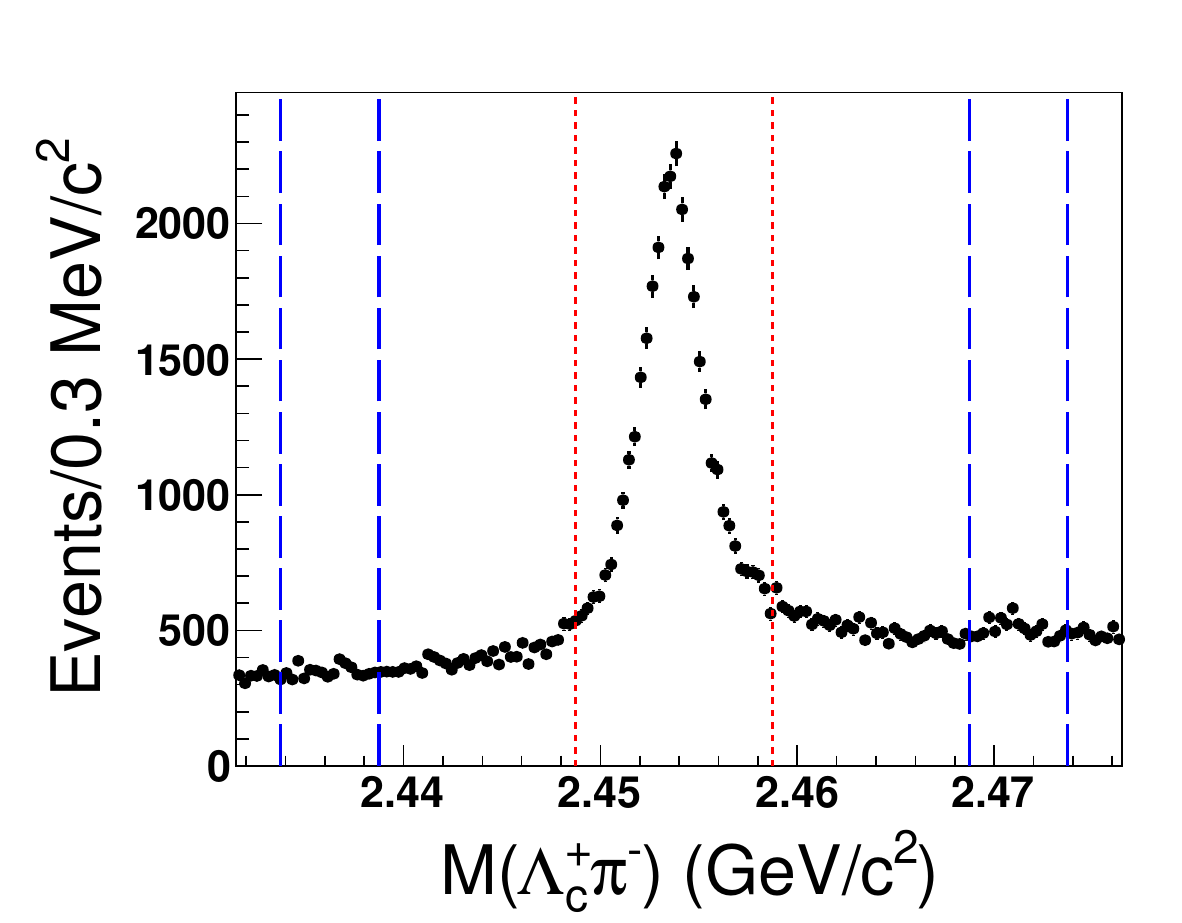}
        \put(-169,143){\bf (a)}
        \quad 
        \includegraphics[width=0.44\textwidth]{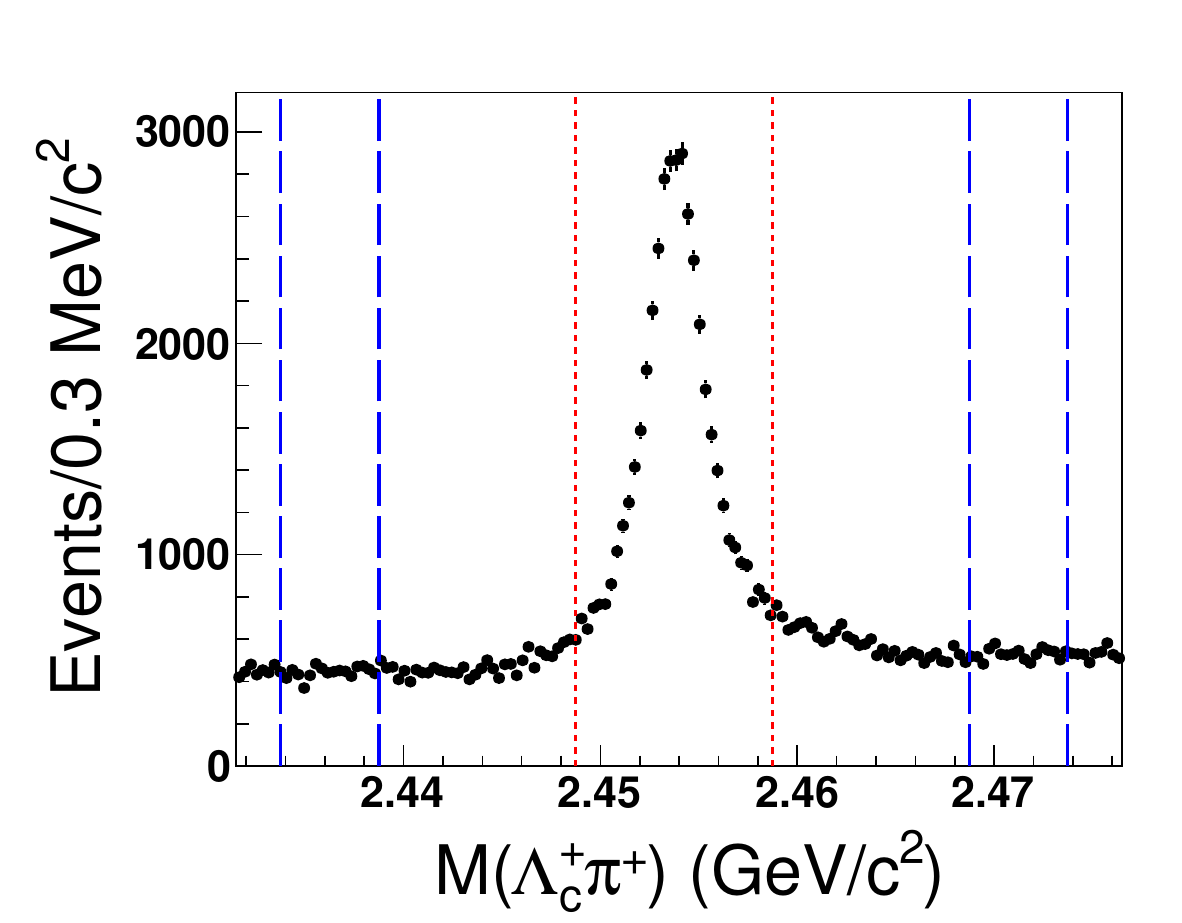}
        \put(-169,143){\bf (b)}
        \caption{Invariant-mass distributions of (a) $\lamc\pi^-$ and (b) $\lamc\pi^+$. Regions between the two red dotted lines are the $\Sigma_c(2455)$ signal regions, and regions between the two blue long-dashed lines are the $\Sigma_c(2455)$ sideband regions.} 
        \label{fig-data-msgm}
        \end{center}
    \end{figure}

     Candidates for $K_S^0 \to \pi^+\pi^-$ are reconstructed from pairs of oppositely charged particles taken as pions, using an artificial neural network~\cite{ks-nn1,ks-nn2}. We require $|M(\pi^+\pi^-)-m(K_S^0)|<10$ MeV/$c^2$, where $m(K_S^0)$ is the nominal $K_S^0$ mass~\cite{pdg}. The two pions are refitted to have a common vertex and the $\pi^+\pi^-$ invariant mass is constrained to $m(K_S^0)$. The $\lamc$ candidates are reconstructed from $\pkpi$ and $\pks$ modes. The invariant masses of $\pkpi$ and $p K^0_S$ are  required to be within $\pm10$ MeV/$c^2$ of the nominal $\lamc$ mass $m(\lamc)$~\cite{pdg}. For the selected events, we constrain $M(p K^- \pi^+)$ and $M(p K^0_S)$ to $m(\lamc)$ to improve momentum resolution. We reconstruct $\Sigma_c(2455)^{0,++}$ {candidates} by combining a $\lamc$ candidate and a charged pion. The requirements of $|M(\lamc \pi^-)-m(\Sigma_c(2455)^{0})| <5$ MeV/$c^2$ and $|M(\lamc \pi^+)-m(\Sigma_c(2455)^{++})| <5$ MeV/$c^2$, defined as $\Sigma_c(2455)$ signal region (SR), are adopted to select $\Sigma_c(2455)^{0,++}$ candidates, where $m(\Sigma_c(2455)^{0,++})$ is the nominal $\Sigma_c(2455)^{0,++}$ mass~\cite{pdg}. The $\Sigma_c(2455)$ sideband regions (SB) are defined as $|M(\lamc \pi^-)-m(\Sigma_c(2455)^{0})| \in (15,~20)$ MeV/$c^2$ and $|M(\lamc \pi^+)-m(\Sigma_c(2455)^{++})| \in (15,~20)$ MeV/$c^2$.  Figure~\ref{fig-data-msgm} shows the $M(\lamc \pi^-)$ and $M(\lamc \pi^+)$ spectra.

     The $D^0$ candidates are reconstructed from $\konepi$, $\ktripi$, and $\ktwopi$ modes, using almost the same selection criteria as {those used} in the study of {$\Xi_c$ states} in the $\Lambda D$ final state~\cite{belle-lamd}. Candidates satisfying $|M(K^-\pi^+)-m(D^0)| < 14$ MeV/$c^2$, $|M(K^-\pi^+\pi^+\pi^-)-m(D^0)| < 11$ MeV/$c^2$, and $|M(K^-\pi^+\pi^0)-m(D^0)| < 27$ MeV/$c^2$, where $m(D^0)$ is the nominal $D^0$ mass~\cite{pdg}, are selected as $D^0$ candidates. For the $\ktwopi$ mode, the energy of the $\pi^0$ candidate is required to be greater than 0.5 GeV to further suppress combinatorial background events. We also require {the} daughter particles of $D^0$ {to originate} from a common vertex and {we constrain the} invariant {mass of the system of final-state particles} to the nominal $D^0$ mass~\cite{pdg}. The sideband regions of $D^0$ are selected as $|M(K^-\pi^+)-m(D^0)| \in (23,~30)$ MeV/$c^2$, $|M(K^-\pi^+\pi^+\pi^-)-m(D^0)| \in(24,~30)$ MeV/$c^2$, and $|M(K^-\pi^+\pi^0)-m(D^0)| \in (38,~50)$ MeV/$c^2$.

     All the above required mass ranges {around the signal peaks} approximately correspond to $\pm 2.5\sigma$ intervals, where $\sigma$ is the standard deviation of the peak distributions. Finally, we reconstruct $\Lambda_c^{*+}$ candidates {from} $\lceta$, $\pdz$, or $\sgmpi$. We define the scaled momentum $x_p$ of $\Lambda_c^{*+}$ as $x_p = p^{*}/\sqrt{s/4-M^2}$~\cite{light-speed} with $p^*$ being the momentum of $\Lambda_c^{*+}$ in the $e^+e^-$ center-of-mass frame, $s$ the square of the center-of-mass energy, and $M$ the mass of $\Lambda_c^{*+}$. We require $x_p>0.7$ to suppress the combinatorial background, especially from $B$-meson decays. There are 5.8\% of the $\Lambda_c^{*+} \to \lceta$ events that have multiple candidates and only the candidate with the smallest $\chi^2(\lamc)+\chi^2(\eta)$ is preserved, where $\chi^2$ stands for the quality of the mass-constrained fit. 

\section{\boldmath search for charmed baryons in the \texorpdfstring{$\lceta$}{Lg} system}
\begin{figure}[htbp]
        \begin{center}
        \includegraphics[width=0.44\textwidth]{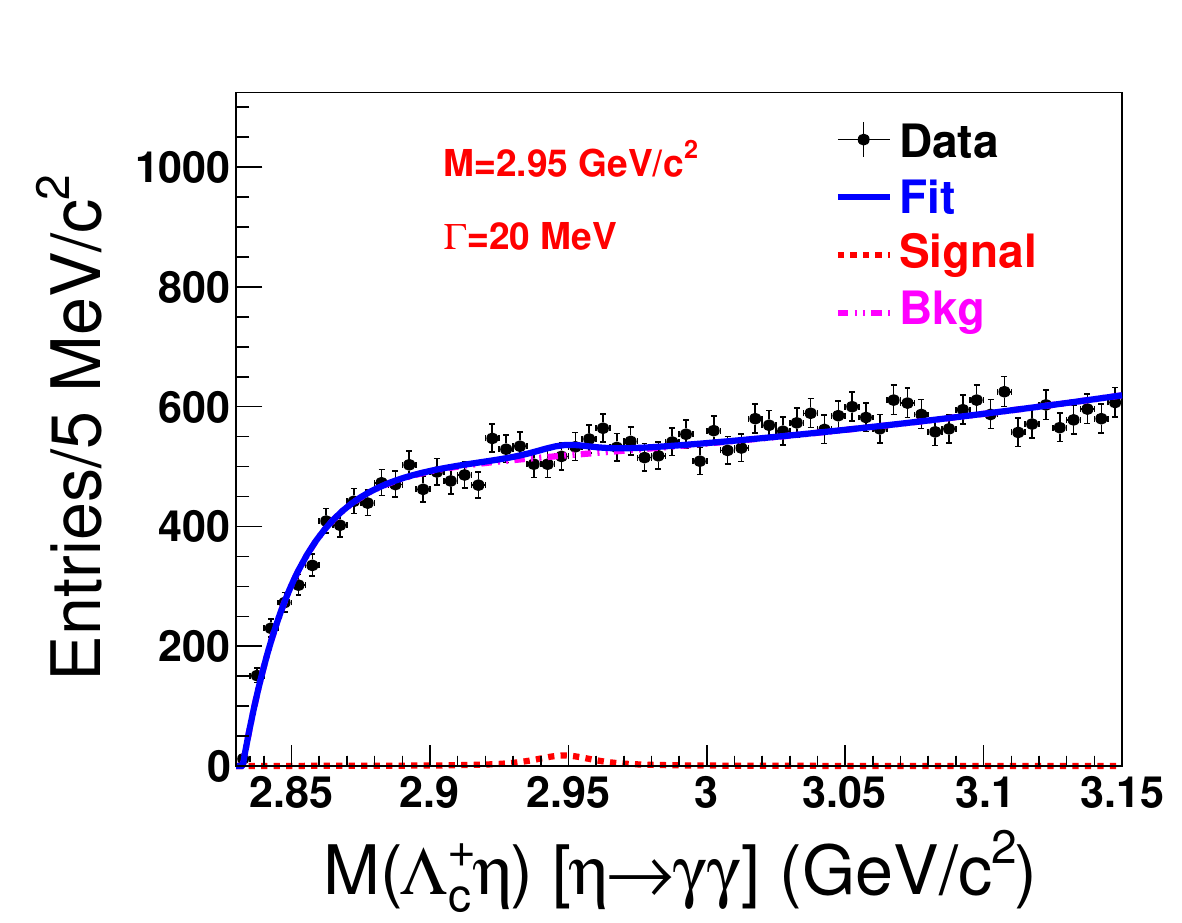}
        \put(-165,143){\bf (a)}
        \quad 
        \includegraphics[width=0.44\textwidth]{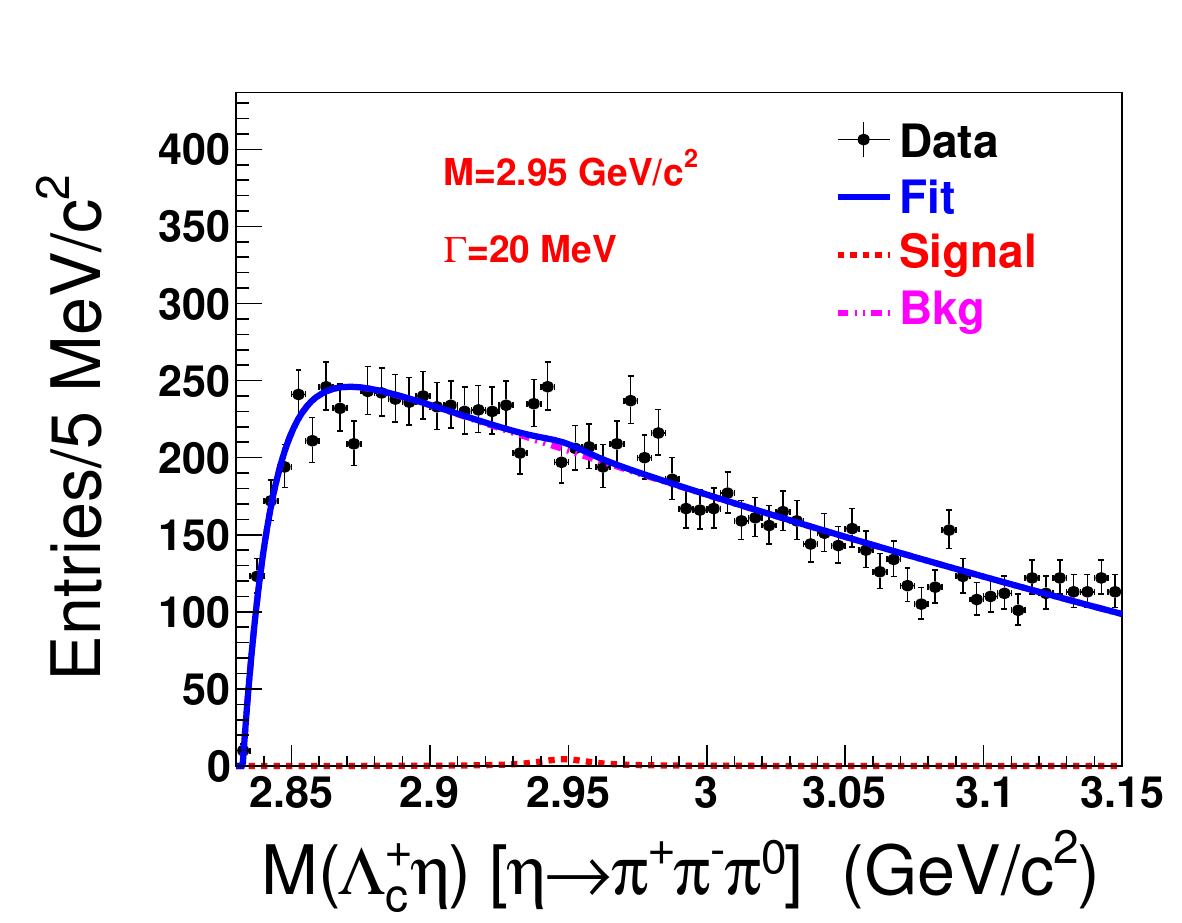}
        \put(-165,143){\bf (b)}
        \caption{An example of the simultaneous fit to the $M(\lceta)$ spectra with (a) $\eta \to \GG$ and (b) $\eta \to \tripi$ (points with statistical uncertainties). The blue solid curves represent the result of the fit. The red dashed curves represent a hypothetical excited $\Lambda_c$ state with $M=2.95$ GeV/$c^2$ and $\Gamma=20$ MeV, while the magenta dash-dotted curves represent the background.} 
        \label{fig-data-lceta}
        \end{center}
    \end{figure}

      Figure~\ref{fig-data-lceta} shows the invariant-mass distributions of $\lceta$ with $\eta \to \GG$ and $\eta \to \tripi$, obtained after applying the selection criteria discussed in Sec.~\ref{sec:selection}. 

     A series of simultaneous binned maximum-likelihood fits to the measured mass spectra of $\lceta$ with $\eta \to \GG$ and $\eta \to \tripi$ are performed. The probability density function (PDF) for the signal component is represented by a constant-width relativistic Breit-Wigner (BW) function convolved with a Gaussian distribution to take the intrinsic mass resolution into account, where the intrinsic mass resolution is determined from signal MC samples. 
     The background component is parametrized by a threshold function of the form 
     \begin{equation}
     \label{eq:lceta-bkg}
     \textstyle
     (1-e^{-\frac{M-M_{\rm th}}{\mu}})(\frac{M}{M_{\rm th}})^a+b(\frac{M}{M_{\rm th}}-1.0)+c(\frac{M}{M_{\rm th}}-1.0)^2.
     \end{equation}
     Here, $M$ represents the $\lceta$ mass, $M_{\rm th}$ stands for the threshold mass, which is fixed to the sum of the nominal $\lamc$ and $\eta$ masses~\cite{pdg}, and $a$, $b$, $c$, and $\mu$ are free parameters. The fit is performed with a 0.5 MeV/$c^2$ bin width, while the histograms in Fig.~\ref{fig-data-lceta} are shown with merged bins of 5 MeV/$c^2$.

     We constrain the signal mass and width to be identical for the two $\eta$ decay modes, and fix the signal mass to be a value from 2.850 to 3.125 GeV/$c^2$ in steps of 25 MeV/$c^2$. The signal width is fixed to 0, 10, 20, 30, and 40 MeV for each fixed mass. The ratio of signal yields for the $\eta \to \GG$ mode relative to those for the $\eta \to \tripi$ mode is calculated by the product of detection efficiencies and the known $\eta$ branching fractions~\cite{pdg}, and is fixed in the fit. This ratio decreases from 5.4 to 2.7 as the signal mass increases from 2.850 to 3.125 GeV/$c^2$.
     Figure~\ref{fig-data-lceta} shows an example of the fit with the signal mass and width fixed to 2.95 GeV/$c^2$ and 20 MeV, respectively.

\begin{figure}[htbp]
        \begin{center}
        \includegraphics[width=0.45\textwidth]{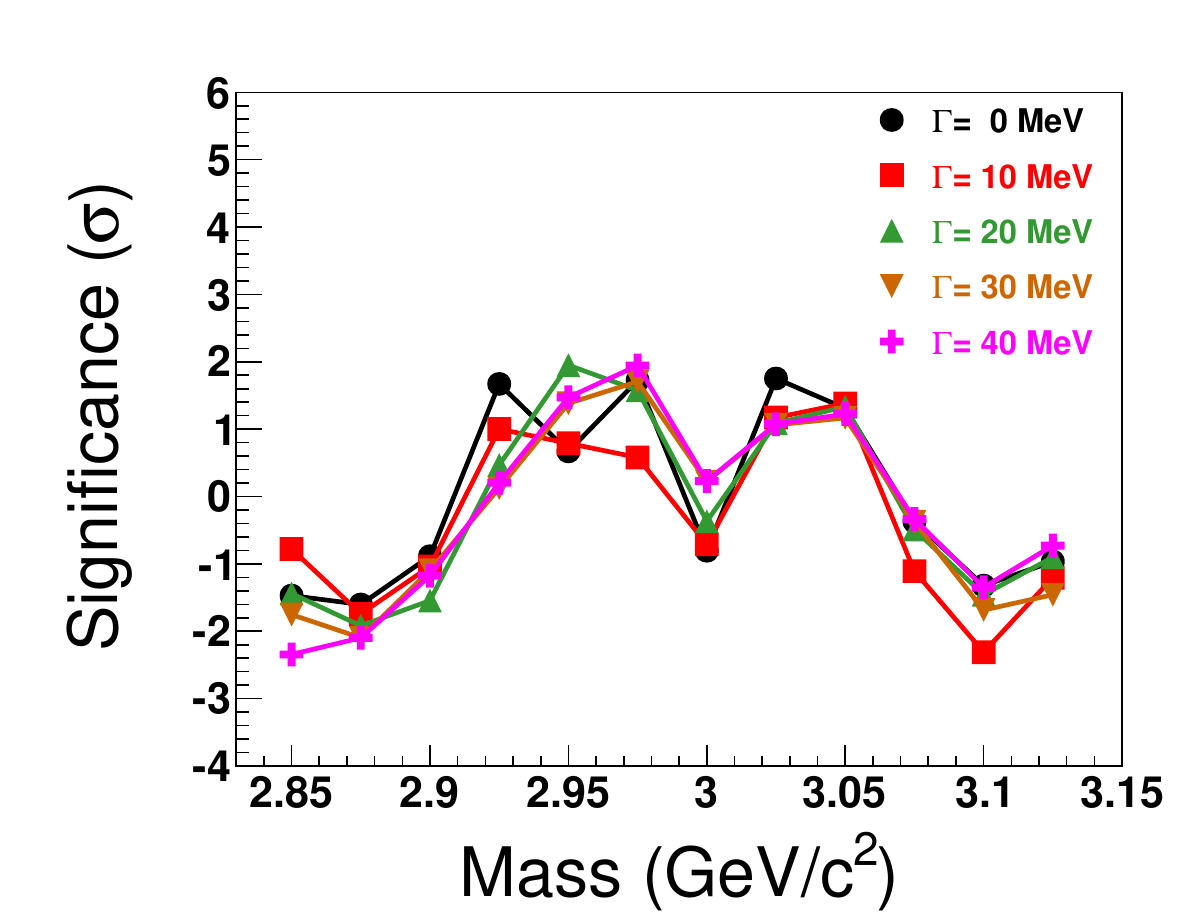}
        \caption{
        Statistical significances for the various excited $\Lambda_c^+$ hypotheses considered.}
        \label{fig-data-significance}
        \end{center}
    \end{figure}

     The statistical significance of the signal is calculated using the log-likelihood difference $-2\text{ln}(\mathcal{L}_0/\mathcal{L}_\text{max})$, where $\mathcal{L}_\text{max}$ and $\mathcal{L}_0$ are the likelihood values of the simultaneous fit to the $\lceta$ mass spectra with and without including the signal PDF, respectively. The significance is negative when the number of fitted signal events is itself negative. The set of statistical significances obtained from the fits is shown in Fig.~\ref{fig-data-significance}. Significances are below $3\sigma$ for all assumed excited $\Lambda_c^+$ hypotheses. We hence find no significant $\Lambda_c^{*+}$ signals in the $\lceta$ invariant-mass spectra.  
      
\section{\boldmath relative Branching fraction}
\begin{figure*}[htbp]
        \begin{center}
        \includegraphics[width=8.9cm]{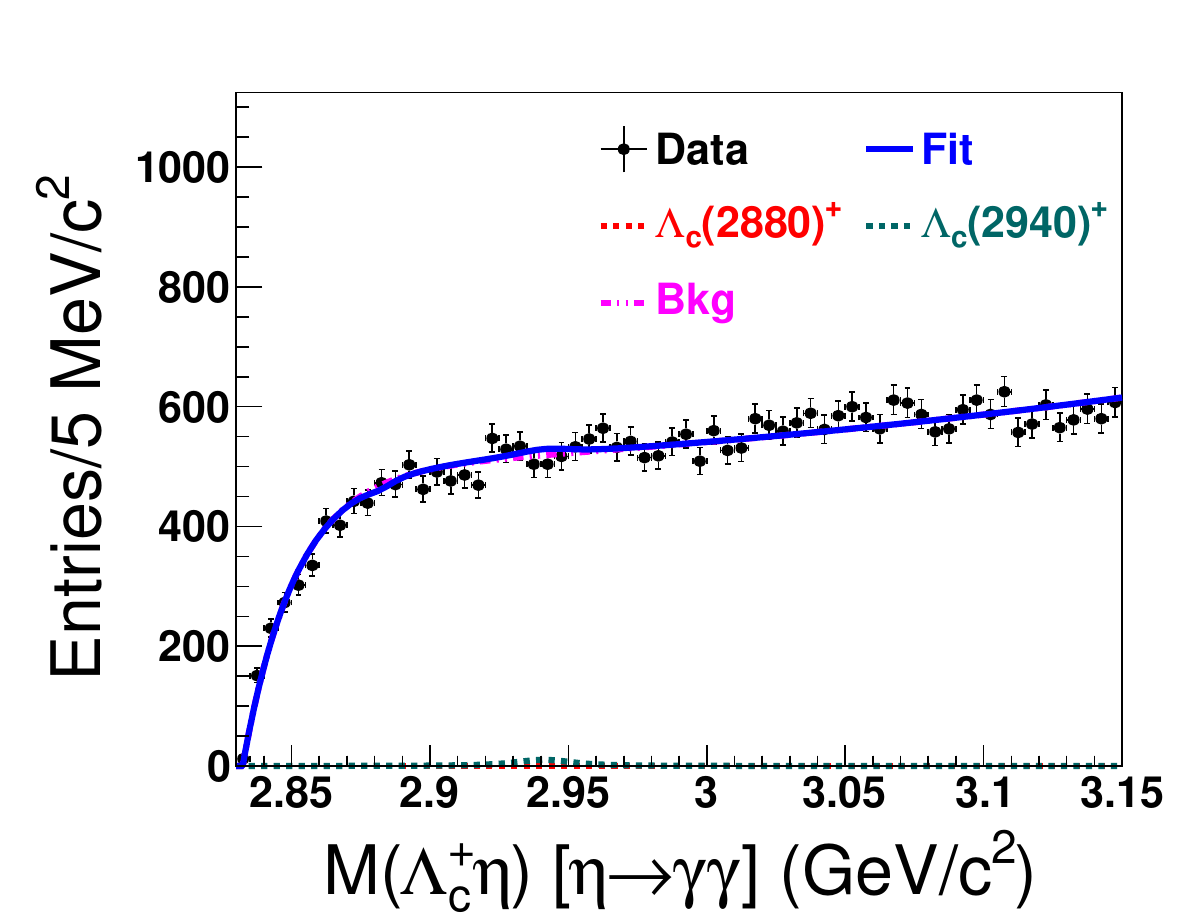}
        \includegraphics[width=8.9cm]{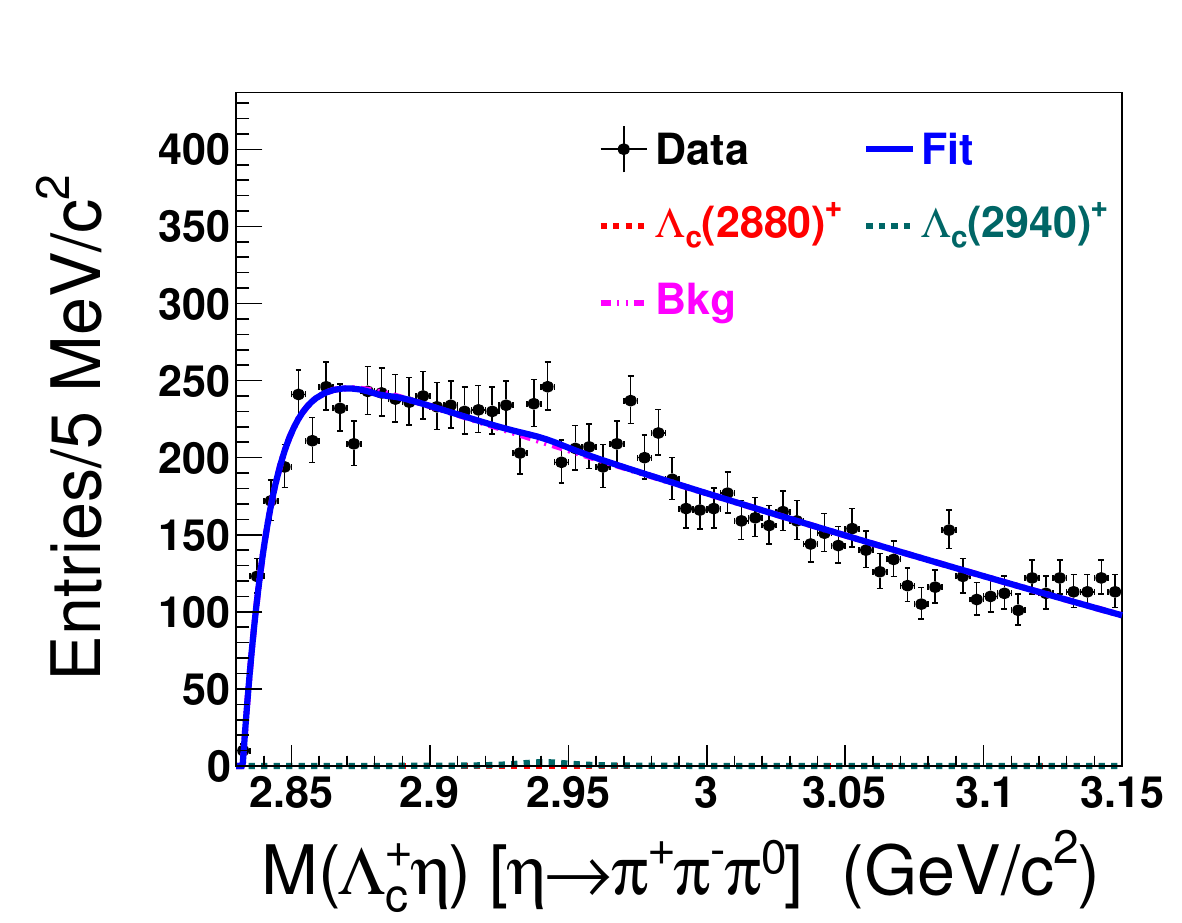}
        \put(-451,165){\bf (a)} 
        \put(-195,165){\bf (b)}
        \quad
        \includegraphics[width=8.9cm]{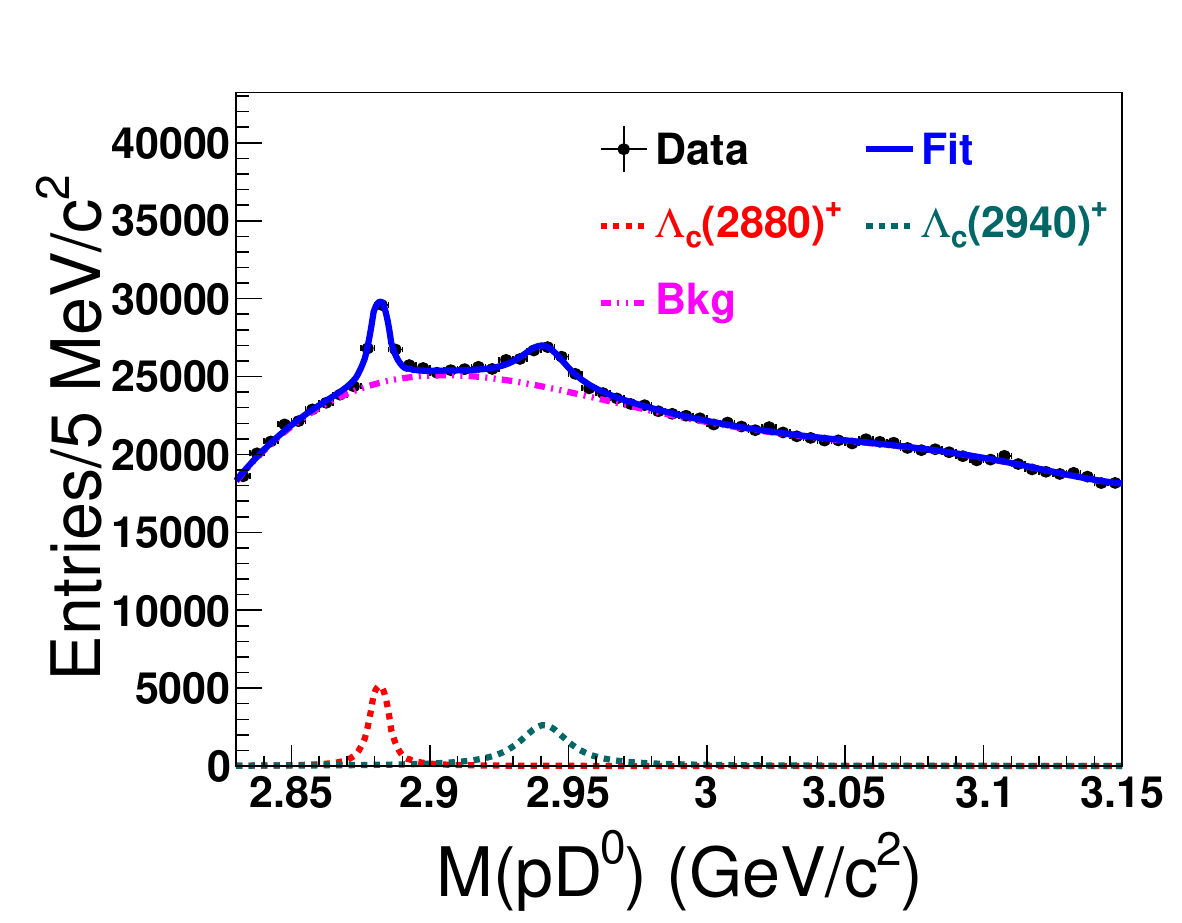}
        \includegraphics[width=8.9cm]{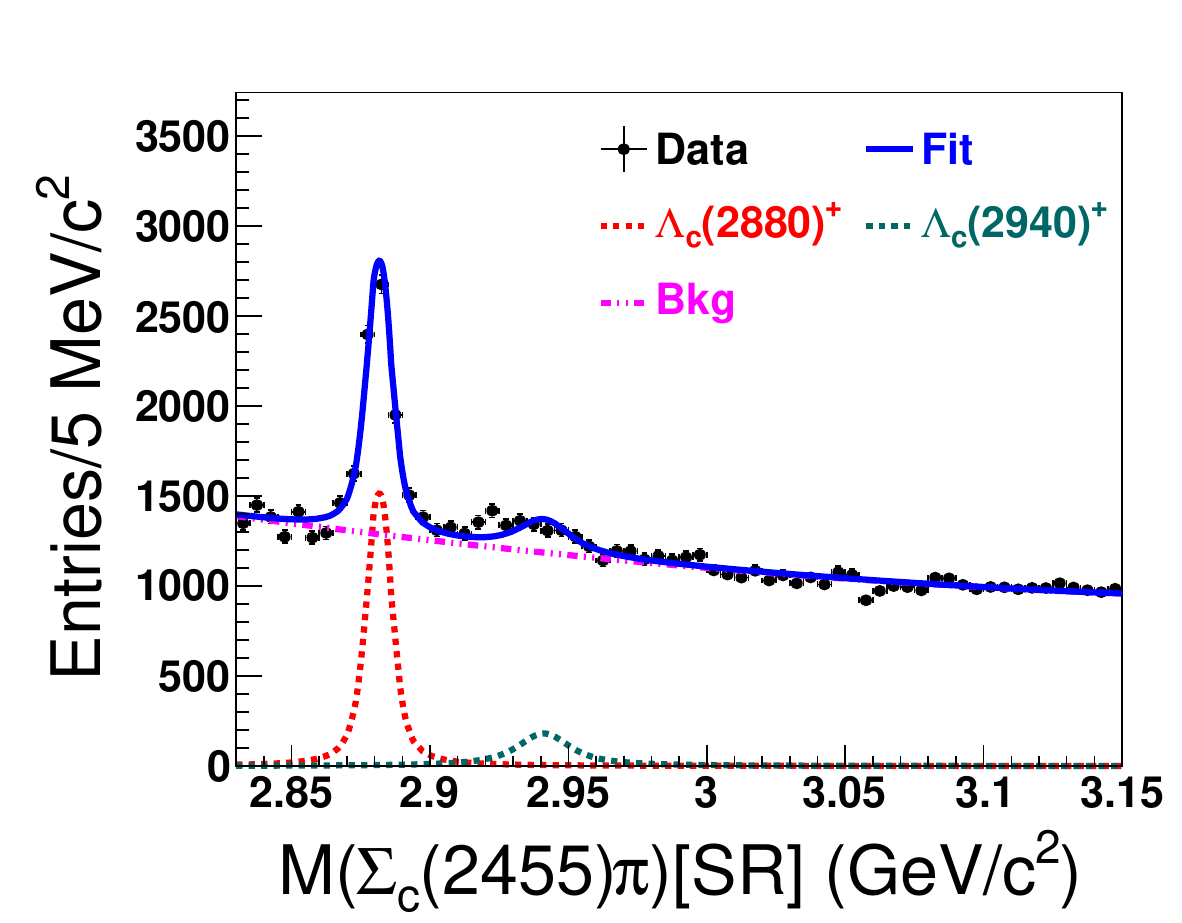}
        \put(-451,165){\bf (c)}
        \put(-195,165){\bf (d)}
        \quad
        \includegraphics[width=8.9cm]{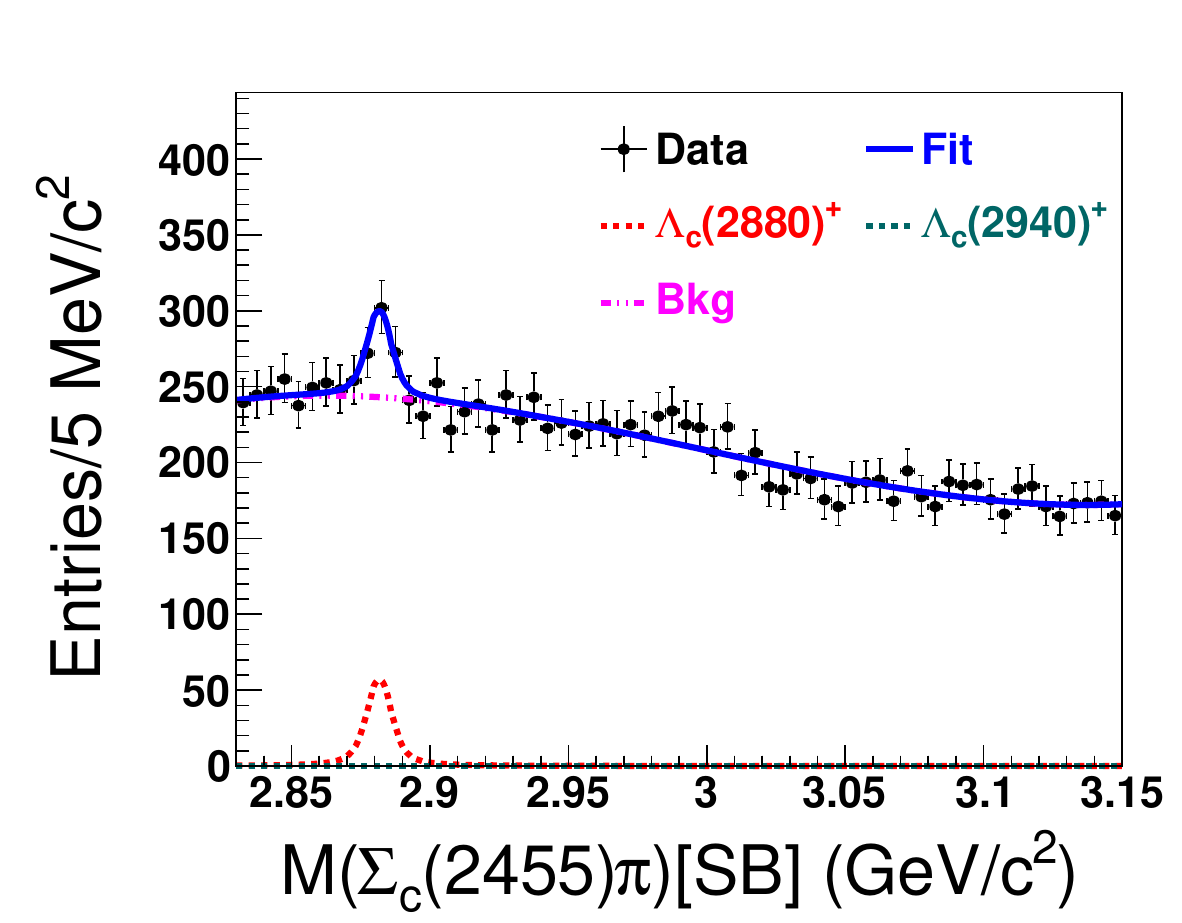}
        \put(-195,165){\bf (e)} 
        \caption{Measured invariant-mass distributions of $\lceta$ with $\eta \to \GG$, $\lceta$ with $\eta \to \tripi$, $\pdz$, $\sgmpi$ in the $\Sigma_c(2455)$ signal region, and $\sgmpi$ in the normalized $\Sigma_c(2455)$ sideband regions  (points with statistical uncertainties). The blue solid curve represents the result of a simultaneous fit. The red and dark green dashed curves represent the  contributions of $\Lambda_c(2880)^+$ and $\Lambda_c(2940)^+$, respectively, while the magenta dash-dotted curves represent the background. }
        \label{fig-data-simpdf}
        \end{center}
     \end{figure*}

     Figure~\ref{fig-data-simpdf} shows the invariant-mass distributions of
     $\lceta$ with $\eta \to \GG$, $\lceta$ with $\eta \to \tripi$, $\pdz$, $\sgmpi$ in the $\Sigma_c(2455)$ signal region, and $\sgmpi$ in the normalized $\Sigma_c(2455)$ sideband regions obtained after applying the selection criteria discussed in Sec.~\ref{sec:selection}. The number of normalized $\Sigma_c(2455)$ sideband events equals to the number of background events in the $\Sigma_c(2455)$ signal region. 
     No clear $\Lambda_c(2880)^+$ or $\Lambda_c(2940)^+$ signals are seen in the $\lceta$ invariant-mass spectra [see Figs.~\ref{fig-data-simpdf}(a) and \ref{fig-data-simpdf}(b)]. By contrast, we observe significant $\Lambda_c(2880)^+$ and $\Lambda_c(2940)^+$ signals in the $M(\pdz)$ distribution [see Fig.~\ref{fig-data-simpdf}(c)]. We do not observe any such peaks in the $\pdz$ mass spectra from generic MC events~\cite{topoana}, nor in the data from the $D^0$ mass sideband regions. Clear $\Lambda_c(2880)^+$ and $\Lambda_c(2940)^+$ signals are also seen in the $M(\sgmpi)$ [SR] distribution [see Fig.~\ref{fig-data-simpdf}(d)].  A $\Lambda_c(2880)^+$ signal is seen in $\Sigma_c(2455)$ sideband events [see Fig.~\ref{fig-data-simpdf}(e)]. This is expected and is due to the $\Lambda_c(2880)^+ \to \lamc\pi^+\pi^-$ non-resonant decay, which in this analysis we consider to be a peaking background to our $\sgmpi$ mode. The contribution of this background is estimated from $\Sigma_c(2455)$ sideband events.

     The yields of $\Lambda_c(2880)^+$ and $\Lambda_c(2940)^+$ are obtained from a simultaneous binned maximum-likelihood fit to the measured mass spectra of $\lceta$ with $\eta \to \GG$, $\lceta$ with $\eta \to \tripi$, $\pdz$, $\sgmpi$ in the $\Sigma_c(2455)$ signal region, and $\sgmpi$ in the normalized $\Sigma_c(2455)$ sideband regions.  In this fit, the  masses of $\Lambda_c(2880)^+$ and $\Lambda_c(2940)^+$ are free parameters,  constrained to be identical for all decay modes, and the widths are fixed to world-average values~\cite{pdg}. The fit is performed with a 0.5 MeV/$c^2$ bin width, while histograms in Fig.~\ref{fig-data-simpdf} are shown with merged bins of 5 MeV/$c^2$. The fitted mass range starts at 2.83 GeV/$c^2$ because of the $\lceta$ threshold. Lower $\Sigma_c(2455)\pi$ and $\Lambda_c^+ \pi^+\pi^-$ masses are populated by the tail of the $\Lambda_c(2765)^+$~\cite{intro-lcpipi} and have been studied already in Ref.~\cite{intro-sgmpi}.

     The PDF for the $\Lambda_c(2880)^+$ or $\Lambda_c(2940)^+$ signal component is represented by a relativistic BW function with mass-dependent width that is convolved with a Gaussian distribution to take the intrinsic mass resolution into account. We determine the intrinsic mass resolution based on signal MC samples. As the most plausible $J^P$ assignment for $\Lambda_c(2880)^+$ is $5/2^+$ and that for $\Lambda_c(2940)^+$ is $3/2^-$, the two signals are parametrized by an $F$-wave BW function and a $D$-wave BW function, respectively, using Blatt-Weisskopf form factors~\cite{blatt}. 
     The background component in the $M(\lceta)$ spectra for $\eta \to \GG$ and $\eta \to \tripi$ is parametrized by the function in Eq.~\ref{eq:lceta-bkg}.  A sixth-order Chebychev polynomial is used to model the combinatorial background component in the $M(p D^0)$ spectrum. The background components in the $M(\sgmpi)$ [SR] and $M(\sgmpi)$ [SB] spectra are parametrized by second-order Chebychev polynomials.

     Since no $\Lambda_c(2880)^+$ and $\Lambda_c(2940)^+$ signals are observed in the $\lceta$ mass spectra, we constrain the ratios of yields of \(\Lambda_c(2880)^+\) and \(\Lambda_c(2940)^+\) in the \(\Lambda_c^+\eta\) decay channel with \(\eta\to\gamma\gamma\) relative to those with $\eta \to \tripi$ by the products of the detection efficiencies and the known $\eta$ branching fractions~\cite{pdg}, resulting in ratio values of 5.1 for $\Lambda_c(2880)^+$ and 4.4 for $\Lambda_c(2940)^+$. The result of the simultaneous fit is shown in Fig.~\ref{fig-data-simpdf}, and the signal yields are summarized in Table~\ref{tab-data-simpdf-yield}. The reduced $\chi^2$ values of the fit are $\chi^2/\rm d.o.f=1.2$ (746.6/635) for $\lceta~(\eta \to \GG)$, 1.5 (937.3/633) for $\lceta~(\eta \to \tripi)$, 1.1 (695.4/629) for $\pdz$, 1.6 (1049.2/634) for $\sgmpi$ in the $\Sigma_c(2455)$ signal region, and 0.5 (297.0/633) for $\sgmpi$ in the normalized $\Sigma_c(2455)$ sideband regions, where $\rm d.o.f$ is the number of degrees of freedom. 

     The possible structures around 2.93, 2.97, and 3.09 GeV/$c^2$ in Fig.~\ref{fig-data-simpdf}(b), which make large contributions to the $\chi^2$, are regarded as statistical fluctuations, since no corresponding structures are found in Fig.~\ref{fig-data-simpdf}(a). The fit to the $M(\sgmpi)$ spectrum in SR, as shown in Fig.~\ref{fig-data-simpdf}(d), shows deviations from the data, and implies that there might be contributions from other states not included in the fit, such as the $\Lambda_c(2910)^+$~\cite{intro-lc2910}. The impact from the $\Lambda_c(2910)^+$ is considered in the systematic uncertainty, as discussed in Section~\ref{sec:syserr}.  
         
     \begin{table}[htbp]
             \centering
             \caption{Signal yields from the simultaneous fit to the mass spectra in Fig.~\ref{fig-data-simpdf}. The uncertainties shown are statistical only.}
             \begin{tabular}{l c@{$\,\pm\,$}c c@{$\,\pm\,$}c } \hline \hline
                 & \multicolumn{2}{c}{$\Lambda_c(2880)^+$} & \multicolumn{2}{c}{$\Lambda_c(2940)^+$} \\ \hline
                $N_{\lceta;~\eta \to \GG}$ & $(-0.2$ & $0.7)\times10^2$ & $(0.7$ & $1.4)\times10^2$ \\
                $N_{\lceta;~\eta \to \tripi}$ & $(-0.04$ & $0.13)\times10^2$ & $(0.16$ & $0.31)\times10^2$ \\
                $N_{\sgmpi}$  & $(4.25$ & $0.15)\times10^3$ & $(1.19$ & $0.19)\times10^3$ \\
                $N_{\rm nonres}$ & $(1.6$ & $0.4)\times10^2$ & $(-0.01$ & $0.03)\times10^2$ \\
                $N_{\pdz}$  &  $(1.17$ & $0.04)\times10^4$ & $(1.64$ & $0.09)\times10^4$ \\ \hline
              \end{tabular} 
              \label{tab-data-simpdf-yield}
     \end{table}

     We measure ratios of branching fractions,
     \begin{equation}
     \small
      R_{\rm X}(2880/2940)\equiv \frac{\mathcal{B}(\Lambda_c(2880/2940)^+ \to \rm X)}{\mathcal{B}(\Lambda_c(2880/2940)^+ \to \sgmpi)},
      \end{equation}
     where $\rm X$ denotes the $\lceta$ or $\pdz$ mode and $\mathcal{B}(\Lambda_c(2880/2940)^+\to \sgmpi)$ is the sum of branching fractions for $\Sigma_c(2455)^0\pi^+$ and $\Sigma_c(2455)^{++}\pi^-$. The two branching fractions are taken as being the same as predicted by isospin symmetry.
     These ratios are obtained from the measured signal yields using
     \begin{equation}
     \left.
     R_{\rm X}=\frac{N_{\rm X}}{\sum_{i}\epsilon^i_{\rm X}\BF^i_{\rm X}}\middle/\frac{N_{\sgmpi}-N_{\rm nonres}}{\sum_j\epsilon^j_{\sgmpi}\BF^j_{\sgmpi}},
     \right.
     \end{equation}
     where the index $i$ represents sub-decay channels for the $\pdz$ or $\lceta$ modes, the index $j$ represents sub-decay channels for the $\sgmpi$ mode, the $\epsilon$ are the detection efficiencies for the decay defined by $\rm X$ and $i$, the $\mathcal{B}$ are the products of branching fractions of intermediate states for the respective decay channels, and the $N$ stand for the signal yields given in Table~\ref{tab-data-simpdf-yield}. The $N_{\rm nonres}$ are the fitted $\Lambda_c(2880)^+$ and $\Lambda_c(2940)^+$ signal yields in the $M(\sgmpi)$ [SB] spectrum. We estimate the detection efficiencies from the MC samples (see Sec.~\ref{sec:sample}), as summarized in Table~\ref{tab-mc-efficiency}, and we calculate the products of known branching fractions using the world-average values from Ref.~\cite{pdg}. 

     \begin{table}[htbp]
             \centering
             \caption{Detection efficiencies from the MC samples for each decay mode.}
             \begin{tabular}{l c c } \hline \hline
                Decays & $\Lambda_c(2880)^+$ & $\Lambda_c(2940)^+$ \\ \hline
                $\lceta~(\pkpi,\GG)$ & 0.0518 & 0.0567 \\
                $\lceta~(\pks,\GG)$  & 0.0578 & 0.0662 \\
                $\lceta~(\pkpi,\tripi)$  & 0.0176 & 0.0224 \\
                $\lceta~(\pks,\tripi)$   & 0.0198 & 0.0241 \\
                $\pdz~(\konepi)$ & 0.273 & 0.289 \\
                $\pdz~(\ktripi)$ & 0.091 & 0.096 \\
                $\pdz~(\ktwopi)$ & 0.176 & 0.185 \\
                $\sgmpi~(\pkpi)$ & 0.137 & 0.144 \\
                $\sgmpi~(\pks)$ & 0.162 & 0.170 \\ \hline
              \end{tabular} 
              \label{tab-mc-efficiency}
     \end{table}

     We obtain branching-fraction ratios for the $\pdz$ mode of $R_{\pdz}(2880)=0.75 \pm 0.03 \pm 0.07$ and $R_{\pdz}(2940)=3.59 \pm 0.21 \pm 0.56$, where the uncertainties are statistical and systematic (see Sec.~\ref{sec:syserr} below for details on the latter ones). 

     Since no significant signals of $\Lambda_c(2880)^+$ and $\Lambda_c(2940)^+$ decaying to $\lceta$ are seen, we set upper limits $R^\text{UL}$ on the branching-fraction ratios at 90\% credibility level (C.L.) by solving the equation
     \begin{equation}
     \frac{\int_0^{R^\text{UL}}\mathcal{L}(R)dR}{\int_0^{+\infty}\mathcal{L}(R)dR}=0.9.
     \end{equation}
     Here, $R$ is the assumed branching-fraction ratio and $\mathcal{L}(R)$ is the corresponding likelihood value obtained from fitting the data. Before integrating, we include the systematic uncertainty $\sigma_{\rm sys}$ described in Sec.~\ref{sec:syserr} by convolving the likelihood function with a Gaussian function whose width is equal to $\sigma_{\rm sys}$. We obtain upper limits on the branching-fraction ratios for the $\lceta$ mode at 90\% C.L. of $R_{\lceta}(2880)<0.13$ and $R_{\lceta}(2940)<1.11$.

\section{\boldmath Systematic Uncertainties}
\label{sec:syserr}
     The systematic uncertainties of the branching-fraction ratios are listed in Table~\ref{tab-data-syserr}. We include uncertainties from the particle identification (PID) efficiency, the tracking efficiency, the $K_S^0$, $\pi^0$, $\eta$, and $\Sigma_c(2455)$  selection efficiency, the branching fractions of intermediate states, the finite MC sample size, the mass resolution, the signal yield extraction, the fit bias, and the impact of $\Lambda_c(2910)^+$ in the $M(\sgmpi)$ spectra.

     \begin{table*}[htbp]
             \centering
             \caption{Relative systematic uncertainties (in \%) for the branching-fraction ratios. \label{tab-data-syserr}}
             \setlength{\tabcolsep}{8.mm}{
             \begin{tabular}{l  c c c c} \hline \hline
                Sources & $R_{\pdz}(2880)$ & $R_{\pdz}(2940)$ & $R_{\lceta}(2880)$  & $R_{\lceta}(2940)$  \\ \hline
                PID efficiency      & 4.5 & 4.5 & 3.5 & 3.5 \\ 
                Tracking efficiency & 0.5 & 0.5 & 0.6 & 0.6 \\
                $K_S^0$ selection & 0.5 & 0.5 & $\cdot\cdot\cdot$ & $\cdot\cdot\cdot$ \\
                $\pi^0$ selection & 0.9 & 0.9 & $\cdot\cdot\cdot$ & $\cdot\cdot\cdot$ \\
                $\eta$  selection & $\cdot\cdot\cdot$ & $\cdot\cdot\cdot$ & 3.0 & 3.0 \\
                $\Sigma_c(2455)$ selection & 1.6 & 2.3 & 1.6 & 2.3 \\
                Branching fraction & 5.1 & 5.1 & 0.6 & 0.6 \\
                MC sample size & 1.3 & 1.2 & 2.2 & 2.2 \\
                Signal yield extraction & 6.2 & 12.5 & 11.2 & 13.5 \\
                Fit bias & 1.0 & 2.3 & 4.1 & 9.6 \\ 
                $\Lambda_c(2910)^+$ & 1.5 & 5.7 & 3.9 & 8.2 \\ \hline
                Total & 9.7 & 15.7 & 13.7 & 19.3 \\ \hline \hline
              \end{tabular} }
    \end{table*}

     The uncertainty from the charged kaon and pion identification efficiency is estimated using $D^{*+}\to D^0\pi^+$ with $D^0 \to K^-\pi^+$, and that of the proton identification efficiency is estimated using $\Lambda^0 \to p\pi^-$. The signal selection efficiency is corrected by a factor derived from the ratio of identification efficiencies of data and MC. We treat the statistical uncertainty of these correction factor as the systematic uncertainty. As the number of kaons, pions, and protons depends on the analyzed sub-decay channels, the uncertainty is calculated by weighting with the product of known branching fractions and detection efficiencies. The uncertainty associated with the PID efficiencies for the $\lamc$ daughters cancels for $R_{\lceta}$. 
 
     The uncertainty of the tracking efficiency is assigned to be 0.35\%  per track, which is estimated using partially reconstructed $D^{*+} \to D^0\pi^+$ with $D^0 \to \pi^+\pi^-K_S^0$. Considering that the number of tracks differs in each sub-decay channel, a weighted uncertainty is calculated. 

     We evaluate the uncertainty coming from the $K_S^0$ selection efficiency using $D^* \to \pi D^0$ with $D^0 \to K_S^0 \pi^0$. The ratio of efficiencies for data and MC is $(98.57\pm0.40)$\%, from which we quote the systematic uncertainty of 
     1.8\%. Since the $K_S^0$ only appears in $\lamc \to \pks$, a weighted uncertainty is calculated.

     Systematic uncertainties due to $\pi^0$ and $\eta$ selection efficiencies are both estimated to be 3\%~\cite{pi0PID}. For $R_{\pdz}$, a weighted uncertainty is calculated since the $\pi^0$ only appears in $D^0 \to K^-\pi^+\pi^0$.

     The ratio of efficiencies of the $M(\Lambda_c^{+}\pi^{\pm})$ requirement for data and MC simulation is $0.968 \pm 0.016$ for $\Lambda_c(2880)^+$, and $0.965\pm0.022$ for $\Lambda_c(2940)^+$. We correct detection efficiencies for the $\sgmpi$ mode with these ratios and take relative uncertainties of ratios as systematic uncertainties, which are 1.6\% for $\Lambda_c(2880)^+$ and 2.3\% for $\Lambda_c(2940)^+$. 

     The systematic uncertainty from the branching fractions of intermediate states are taken from world-average values in Ref.~\cite{pdg}. The uncertainty related to the branching fractions of the $\lamc$ and $K^0_S$ decays cancels for $R_{\lceta}$.

     The relative uncertainty due to the finite MC sample size is estimated as $\sqrt{(1-\epsilon)/(\epsilon N_{\rm gen})}$, where $\epsilon$ is the detection efficiency and $N_{\rm gen}$ is the number of generated signal MC events.

     The systematic uncertainty due to the fixed mass resolution is estimated by fitting the $M(\pdz)$ spectrum with the mass resolution floated. The relative difference on the ratio of the branching fractions between the one from the floated mass resolution and the nominal value is taken as the uncertainty, which is found to be negligible.

     The systematic uncertainty from the signal yield extraction consists of  uncertainties from the widths of $\Lambda_c(2880)^+$ and $\Lambda_c(2940)^+$, the background PDF, the fit interval, and the bin width. To estimate the uncertainty introduced by fixing the widths of $\Lambda_c(2880)^+$ and $\Lambda_c(2940)^+$, we vary the signal width by $\pm1\sigma$~\cite{pdg} and take the maximum difference from the nominal result as the systematic uncertainty. The uncertainty arising from the background PDF is determined by varying the order of the Chebychev polynomial. The uncertainty from the fit range is estimated by enlarging the fitted interval from (2.83, 3.15) GeV/$c^2$ to (2.81, 3.17) GeV/$c^2$. We evaluate the uncertainty due to the bin width by varying it from 0.5 MeV/$c^2$ to 0.6 MeV/$c^2$ or 0.4 MeV/$c^2$, and taking the maximum difference as the systematic uncertainty. Of the above uncertainties, the ones due to the signal width and fit interval are dominant and those from the background PDF and the bin width are small.
     The various contributions to the yield uncertainty are added in quadrature. 

     The systematic uncertainty introduced by the fit bias is assessed by constructing an ensemble of MC pseudo-experiments. For each pseudo-experiment, we generate  simulated MC samples, by randomly drawing signal and background MC events from the signal and background PDFs, respectively, based on the extracted numbers of corresponding events in data. The sizes of these MC samples are hence identical to the extracted yields for each decay mode. We repeat the above procedure 1000 times and then perform simultaneous binned  maximum-likelihood fits on these 1000 independent pseudo-experiments. This yields Gaussian-shaped distributions of the $R_{\rm X}$ values. The relative differences between the mean values of these distributions and the nominal $R_{\rm X}$ values are assigned as the systematic uncertainties. 

     The systematic uncertainty from the possible $\Lambda_c(2910)^+$ signal in the $M(\sgmpi)$ spectrum around 2.90-2.95 GeV/$c^2$ is estimated by adding an additional state, the $\Lambda_c(2910)^+$, in the fit to the $M(\sgmpi)$ spectra. The relative difference of the branching-fraction ratio is taken as the systematic uncertainty.

     Summing all the uncertainty terms discussed above in quadrature gives the total systematic uncertainty quoted in Table~\ref{tab-data-syserr}.

\section{\boldmath summary}
     In this paper, we report on a search for excited singly-charmed baryons performed for the first time in the $\lceta$ mass spectra in a range from 2.83 to 3.15 GeV/$c^2$ based on the full Belle dataset corresponding to an integrated luminosity of 980 $\rm fb^{-1}$. No significant excess is found in the $M(\lceta)$ spectrum. This is in contrast to excited hyperons, where resonances decaying into $\Lambda \eta$ have been observed.

     Clear $\Lambda_c(2880)^+$ and $\Lambda_c(2940)^+$ signals are observed in the $\pdz$ mass spectrum. The first measurements of branching-fraction ratios of $\Lambda_c(2880)^+$ and $\Lambda_c(2940)^+$ decaying to $\lceta$ and $\pdz$ relative to $\sgmpi$ are performed. We measure
     \begin{align*}
       R_{\pdz}(2880)&=0.75 \pm 0.03 \pm 0.07,\\
       R_{\pdz}(2940)&=3.59 \pm 0.21 \pm 0.56,
     \end{align*}
     where the first uncertainty is statistical and the second one systematic. We also determine upper limits for $R_{\lceta}$ at 90\% C.L. of
     \begin{align*}
       R_{\lceta}(2880)&<0.13,\\
       R_{\lceta}(2940)&<1.11.
     \end{align*} 
     It is predicted in Ref.~\cite{intro-qm} that the $\rho$-mode  states decay primarily into a light meson and a heavy baryon,
     whereas the  $\lambda$-mode  states decay predominantly into a light baryon and a heavy meson. Therefore, when interpreted
     using this model, our branching-ratio measurements suggest that the $\Lambda_c(2880)^+$ is a $\rho$-mode excited state
     while the $\Lambda_c(2940)^+$ is a $\lambda$-mode excited state.

\section*{ACKNOWLEDGMENTS}

This work, based on data collected using the Belle detector, which was
operated until June 2010, was supported by 
the Ministry of Education, Culture, Sports, Science, and
Technology (MEXT) of Japan, the Japan Society for the 
Promotion of Science (JSPS), and the Tau-Lepton Physics 
Research Center of Nagoya University; 
the Australian Research Council including grants
DP210101900, 
DP210102831, 
DE220100462, 
LE210100098, 
LE230100085; 
Austrian Federal Ministry of Education, Science and Research and
Austrian Science Fund (FWF) No.~P~31361-N36;
National Key R\&D Program of China under Contract No.~2022YFA1601903,
National Natural Science Foundation of China and research grants
No.~11575017,
No.~11761141009, 
No.~11705209, 
No.~11975076, 
No.~12135005, 
No.~12150004, 
No.~12161141008, 
and
No.~12175041, 
and Shandong Provincial Natural Science Foundation Project ZR2022JQ02;
the Czech Science Foundation Grant No. 22-18469S;
Horizon 2020 ERC Advanced Grant No.~884719 and ERC Starting Grant No.~947006 ``InterLeptons'' (European Union);
the Carl Zeiss Foundation, the Deutsche Forschungsgemeinschaft, the
Excellence Cluster Universe, and the VolkswagenStiftung;
the Department of Atomic Energy (Project Identification No. RTI 4002), the Department of Science and Technology of India,
and the UPES (India) SEED finding programs Nos. UPES/R\&D-SEED-INFRA/17052023/01 and UPES/R\&D-SOE/20062022/06; 
the Istituto Nazionale di Fisica Nucleare of Italy; 
National Research Foundation (NRF) of Korea Grant
Nos.~2016R1\-D1A1B\-02012900, 2018R1\-A2B\-3003643,
2018R1\-A6A1A\-06024970, RS\-2022\-00197659,
2019R1\-I1A3A\-01058933, 2021R1\-A6A1A\-03043957,
2021R1\-F1A\-1060423, 2021R1\-F1A\-1064008, 2022R1\-A2C\-1003993;
Radiation Science Research Institute, Foreign Large-size Research Facility Application Supporting project, the Global Science Experimental Data Hub Center of the Korea Institute of Science and Technology Information and KREONET/GLORIAD;
the Polish Ministry of Science and Higher Education and 
the National Science Center;
the Ministry of Science and Higher Education of the Russian Federation
and the HSE University Basic Research Program, Moscow; 
University of Tabuk research grants
S-1440-0321, S-0256-1438, and S-0280-1439 (Saudi Arabia);
the Slovenian Research Agency Grant Nos. J1-9124 and P1-0135;
Ikerbasque, Basque Foundation for Science, and the State Agency for Research
of the Spanish Ministry of Science and Innovation through Grant No. PID2022-136510NB-C33 (Spain);
the Swiss National Science Foundation; 
the Ministry of Education and the National Science and Technology Council of Taiwan;
and the United States Department of Energy and the National Science Foundation.
These acknowledgements are not to be interpreted as an endorsement of any
statement made by any of our institutes, funding agencies, governments, or
their representatives.
We thank the KEKB group for the excellent operation of the
accelerator; the KEK cryogenics group for the efficient
operation of the solenoid; and the KEK computer group and the Pacific Northwest National
Laboratory (PNNL) Environmental Molecular Sciences Laboratory (EMSL)
computing group for strong computing support; and the National
Institute of Informatics, and Science Information NETwork 6 (SINET6) for
valuable network support.


\end{document}

%% file: pub676-orcid.tex
\noaffiliation
  \author{S.~X.~Li\,\orcidlink{0000-0003-4669-1495}} 
  \author{C.~P.~Shen\,\orcidlink{0000-0002-9012-4618}} 
  \author{I.~Adachi\,\orcidlink{0000-0003-2287-0173}} 
  \author{J.~K.~Ahn\,\orcidlink{0000-0002-5795-2243}} 
  \author{H.~Aihara\,\orcidlink{0000-0002-1907-5964}} 
  \author{D.~M.~Asner\,\orcidlink{0000-0002-1586-5790}} 
  \author{H.~Atmacan\,\orcidlink{0000-0003-2435-501X}} 
  \author{T.~Aushev\,\orcidlink{0000-0002-6347-7055}} 
  \author{R.~Ayad\,\orcidlink{0000-0003-3466-9290}} 
  \author{Sw.~Banerjee\,\orcidlink{0000-0001-8852-2409}} 
% ...
\collaboration{The Belle Collaboration}

%% file: draft.bbl
\begin{thebibliography}{99}
\bibitem{intro-theo1}
J.~M.~Richard, Phys. Rep. {\bf 212}, 1 (1992).

\bibitem{intro-theo2}
E.~Klempt and J.~M.~Richard, Rev. Mod. Phys. {\bf 82}, 1095 (2010).

\bibitem{intro-theo3}
V.~Crede and W.~Roberts, Rep. Prog. Phys. {\bf 76}, 076301 (2013).

\bibitem{intro-theo4}
H.~X.~Chen, W.~Chen, X.~Liu, Y.~R.~Liu, and S.~L.~Zhu, Rep. Prog. Phys. {\bf 80}, 076201 (2017).

\bibitem{intro-theo5}
H.~Y.~Cheng, Chin. J. Phys. {\bf 78}, 324 (2022).

\bibitem{intro-qm}
T.~Yoshida, E.~Hiyama, A.~Hosaka, M.~Oka, and K.~Sadato, Phys. Rev. D {\bf 92}, 114029 (2015).


\bibitem{intro-lcpipi}
M.~Artuso {\it et al.} (CLEO collaboration), Phys. Rev. Lett. {\bf 86}, 4479 (2001).

\bibitem{intro-pdz}
B.~Aubert {\it et al.} (BaBar Collaboration), Phys. Rev. Lett. {\bf 98}, 012001 (2007).

\bibitem{intro-sgmpi}
R.~Mizuk {\it et al.} (Belle Collaboration), Phys. Rev. Lett. {\bf 98}, 262001 (2007).

\bibitem{intro-lc2940-jp}
R.~Aaij {\it et al.} (LHCb Collaboration), J. High. Energy. Phys. {\bf 05}, 030 (2017).

\bibitem{intro-ND1}
J.~R.~Zhang, Phys. Rev. D {\bf 89}, 096006 (2014).

\bibitem{intro-ND2}
Y.~Dong, Phys. Rev. D {\bf 81}, 014006 (2010).

\bibitem{intro-ND3}
 P.~G.~Ortega, D.~R.~Entem, and F.~Fernandez,  Phys. Lett. B {\bf 718}, 1381 (2013).

\bibitem{intro-lbd2000}
H. Zhang, J. Tulpan, M. Shrestha, and D. M. Manley, Phys. Rev. C {\bf 88}, 035205 (2013).

\bibitem{intro-ld}
S.~B.~Yang {\it et al.} (Belle Collaboration), Phys. Rev. D {\bf 108}, L031104 (2023). 

\bibitem{intro-kn1}
N.~Kaiser, P.~B.~Siegel, and W.~Weise, Nucl. Phys. A {\bf 594}, 325 (1995).

\bibitem{intro-kn2}
E.~Oset and A.~Ramos,  Nucl. Phys. A {\bf 635}, 99 (1998).

\bibitem{intro-kn3}
J.~A.~Oller and U.~G.~Meißner, Phys. Lett. B {\bf 500}, 263 (2001).

\bibitem{KEKB1}
S.~Kurokawa and E.~Kikutani, Nucl. Instrum. Methods Phys. Res., Sect. A  {\bf 499}, 1 (2003), and other papers included in this volume.

\bibitem{KEKB2}
T.~Abe {\it et al.}, Prog. Theor. Exp. Phys. {\bf 2013}, 03A001 (2013), and references therein.

\bibitem{Belle1}
J.~Brodzicka {\it et al.}, Prog. Theor. Exp. Phys. {\bf 2012}, 04D001 (2012).

\bibitem{Belle2}
A.~Abashian {\it et al.} (Belle Collaboration), Nucl. Instrum. Methods Phys. Res., Sect. A {\bf 479}, 117 (2002).

\bibitem{evtgen}
D.~J.~Lange, Nucl. Instrum. Methods Phys. Res., Sect. A {\bf 462}, 152 (2001).

\bibitem{geant3}
R.~Brun {\it et al.}, GEANT 3: user’s guide Geant 3.10, Geant 3.11, CERN Report No. DD/EE/84-1, 1984.

\bibitem{pythia}
T.~Sj\"{o}strand, S.~Mrenna, and P.~Skands, Comput. Phys. Commun. {\bf 178}, 852 (2008).

\bibitem{pdg}
R.~L.~Workman {\it et al.} (Particle Data Group), Prog. Theor. Exp. Phys. {\bf 2022}, 083C01 (2022) and 2023 update.

\bibitem{phsp}
Charge-conjugate modes are implied throughout this paper unless otherwise stated.

\bibitem{photons}
E.~Barberio and Z.~Was, Comput. Phys. Commun. {\bf 79}, 291 (1994).


\bibitem{pidcode}
E.~Nakano, Nucl. Instrum. Methods Phys. Res., Sect. A {\bf 494}, 402 (2002).

\bibitem{eidcode}
K.~Hanagaki, H.~Kakuno, H.~Ikeda, T.~Iijima, and T.~Tsukamoto, Nucl. Instrum. Methods Phys. Res., Sect. A {\bf 485}, 490 (2002).

\bibitem{ks-nn1}
M.~Feindt and U.~Kerzel, Nucl. Instrum. Methods Phys. Res., Sect. A {\bf 559}, 190 (2006).

\bibitem{ks-nn2}
H.~Nakano, Ph.D. thesis, Tohoku University, 2014, Chap. 4, http://hdl.handle.net/10097/58814.

\bibitem{belle-lamd}
Y.~Kato {\it et al.} (Belle Collaboration), Phys. Rev. D {\bf 94}, 032002 (2016).

\bibitem{light-speed}
We used units in which the speed of light is $c=1$.

\bibitem{topoana}
X.~Y.~Zhou, S.~X.~Du, G.~Li, and C.~P.~Shen, Comput. Phys. Commun. {\bf 258}, 107540 (2021).

\bibitem{blatt}
S.~Godfey and N.~Isgur, Phys. Rev. D {\bf 32}, 180. (1985).

\bibitem{intro-lc2910}
Y.~B.~Li {\it et al.} (Belle Collaboration), Phys. Rev. Lett. {\bf 130}, 031901 (2023).

\bibitem{pi0PID}
M.~C.~Chang {\it et al.}, Phys. Rev. D {\bf 85}, 091102 (2012).






\end{thebibliography}
